\documentstyle[aps,prd,multicol]{revtex}
\tighten

\def\Lrule{\vspace*{-0.2in}\noindent\vrule width3.4in height.2pt
  depth.2pt \vrule depth0em height.5em}
\def\Rrule{\vspace{-0.1in}\hfill\vrule depth.5em height0pt \vrule
  width3.4in height.2pt depth.2pt\vspace*{-0.1in}}

\begin{document}

\title{Hamiltonian embedding of the massive noncommutative U(1)
theory}

\author{R. Amorim$^a$ and J. Barcelos-Neto$^b$}

\address{\mbox{}\\
Instituto de F\'{\i}sica\\
Universidade Federal do Rio de Janeiro\\
RJ 21945-970 - Caixa Postal 68528 - Brasil}
\date{\today}

\maketitle
\begin{abstract}
\hfill{\small\bf Abstract\hspace*{1.7em}}\hfill\smallskip
\par
\noindent
We show that the massive noncommutative U(1) can be embedded in
a gauge theory by using the BFFT Hamiltonian formalism. By virtue of
the peculiar non-Abelian algebraic structure of the noncommutative
massive $U(1)$ theory, several specific identities involving Moyal
commutators had to be used in order to make the embedding possible.
This leads to an infinite number of steps in the iterative process of
obtaining first-class constraints. We also shown that the involutive
Hamiltonian can be constructed.
\end{abstract}

\smallskip\mbox{}

\begin{multicols}{2}
\section{Introduction}
\renewcommand{\theequation}{1.\arabic{equation}}

The embedding procedure is an interesting mechanism where theories
with second-class constraints \cite{Dirac} are transformed into more
general (gauge) theories in such a way that all constraints become
first-class. Once this is achieved it is possible to exploit the
machinery available for quantizing first-class theories
\cite{Henneaux}. The use of Dirac brackets is completely avoided. The
important point of this mechanism is that the general theory so
obtained contains all the physics of the embedded one.

\medskip
A systematic formalism for embedding was developed by Batalin,
Fradkin, Fradkina, and Tyutin (BFFT) \cite{BFFT1,BFFT2}. A fundamental
characteristic of the method is the extension of the original phase
space by introducing one auxiliary field for each second-class
constraint. The BFFT method is quite elegant and the obtainment of
first-class constraints is done in an iterative way. The first
correction is linear in the auxiliary variables, the second one is
quadratic, and so on. In the case of systems with just linear
constraints, like chiral-bosons \cite{Floreanini}, only linear
corrections are enough to make them first-class \cite{Miao,Barc1}.
Here, we mention that the method is equivalent to express the dynamic
quantities by means of shifted coordinates \cite{Amorim}.

\medskip
However, for systems with nonlinear constraints, the iterative
process may go beyond the first correction. This is a crucial point
because the first iterative step does not give a unique solution and
we do not know {\it a priori} what should be the most convenient
one we have to choose for the second step. There are systems
where this choice can be done in a special way such that the
iterative procedure stops and it is not necessary to go to higher
orders \cite{Banerjee}. There are other systems, like the massive
Yang-Mills theory, where it is feasible to carry out all the steps of
the method \cite{Barc2,Park}.

\medskip
It is important to emphasize that the arbitrariness in choosing some
specific solution is not a blemish of the method. It just tell us that
there can be more than one way of embedding some specific theory. On
the other hand, it is also opportune to say that there are theories,
like the full nonlinear sigma model, where no embedding is possible
\cite{Barc3}.

\medskip
Recently, there have been a great deal of interest in noncommutative
field theories. This has started when it was noted that noncommutative
spaces naturally arise in the study of perturbative string theory in
the presence of $D$-branes with a constant background magnetic field.
The dynamics of the $D$-brane in this limit can be described by a
noncommutative gauge theory \cite{Witten}. Besides their origin in
strings and branes, noncommutative field theories are a very
interesting subject by their own rights. They have been studied
extensively in many branches of field theory. We mention, among
others, $C$, $P$, and $T$ invariance \cite{Sheikh}, axial anomaly
\cite{Ardalan}, noncommutative QED \cite{Dayi}, supersymmetry
\cite{Girotti}, renormalization and the mixing of infrared and
ultraviolet divergencies \cite{Seiberg}, unitarity \cite{Gomis}, and
phenomenology \cite{Arfaei}.

\medskip
To obtain the noncommutative version of a field theory one essentially
replaces the product of fields in the action by the Moyal product:

\begin{equation}
\phi_1(x)\star\phi_2(x)=\exp\,
\left(\frac{i}{2}\theta^{\mu\nu}\partial_\mu^x\partial_\nu^y\right)\,
\phi_1(x)\phi_2(y)\vert_{x=y}
\label{1.1}
\end{equation}

\noindent
where $\theta^{\mu\nu}$ is a real and antisymmetric constant matrix.
It is easily verified that the Moyal product of two fields in the
action is the same as the usual product, provide we discard boundary
terms. In this way, the noncommutativity affects just the vertices.

\medskip
In the present paper, we are going to consider the Hamiltonian
embedding of the noncommutative version of the massive QED. By virtue
of the Moyal commutators, this theory is non-Abelian and has an
interesting mathematical structure that makes it completely different
from the usual non-Abelian Yang-Mills case. The use of the BFFT method
leads to infinite iterative steps. We are going to see that the
solution of the equations corresponding to the initial steps as well
as the inferring of the general solution are only achieved by using
several identities related to the specific algebra of the Moyal space.
We show that the second-class constraints and the Hamiltonian are
transformed to form an involutive system of dynamical quantities that
are defined as series of Moyal commutators among the variables
belonging to the BFFT extended phase space. We also make explicit the
gauge invariance of the first order system so obtained.

\medskip
Our paper is organized as follows. In Sec. II we make a brief
introduction of the massive noncommutative $U(1)$ theory in order to
fix the notation and convention we are going to use through the
paper. The embedding of the constraints is done in Sec. III and the
obtainment of the involutive Hamiltonian, as well as the question of
gauge invariance are considered in Sec. IV. We left Sec. V for some
concluding remarks. We include four appendices to list the most
important identities and present some details of relevant
calculations.

\section{The noncommutative massive U(1) theory}
\renewcommand{\theequation}{2.\arabic{equation}}
\setcounter{equation}{0}

Let us start from the massless case. The corresponding action reads

\begin{eqnarray}
S&=&-\,\frac{1}{4}\int d^4x\,F_{\mu\nu}\star F^{\mu\nu}
\nonumber\\
&=&-\,\frac{1}{4}\int d^4x\,F_{\mu\nu}F^{\mu\nu}
\label{2.1}
\end{eqnarray}

\noindent
However, the stress tensor is defined in terms of the Moyal commutator
of the covariant derivatives, i.e.,

\begin{eqnarray}
ie\,F_{\mu\nu}&=&[D_\mu,D_\nu]
\nonumber\\
&=&D_\mu\star D_\nu-D_\nu\star D_\mu
\label{2.2}
\end{eqnarray}

\noindent
Considering the covariant derivative

\begin{equation}
D_\mu=\partial_\mu-ie\,A_\mu
\label{2.3}
\end{equation}

\noindent
one obtains

\begin{eqnarray}
F_{\mu\nu}&=&\partial_\mu A_\nu-\partial_\nu A_\mu
-ie\,(A_\mu\star A_\nu-A_\nu\star A_\mu)
\nonumber\\
&=&\partial_\mu A_\nu-\partial_\nu A_\mu
-ie\,[A_\mu,A_\nu]
\label{2.4}
\end{eqnarray}

We observe that the noncommutative stress tensor has the same general
form as the corresponding Yang-Mills theory. There, gauge fields
couple to themselves by means of color charges related to  generators
of the $SU(N)$ symmetry group. Here, the gauge field coupling is due
to the mathematical definition of the Moyal product and does not have
the same physical meaning of the color charge coupling. The
non-Abelian character of the two cases are completely different, both
in physical and in mathematical points of view.

\medskip
The gauge transformation of $A_\mu$ can be obtained by considering
that
the gauge transformation of the covariant derivative acting on some
charged field $\psi$ must satisfy the relation

\begin{equation}
(D_\mu\star\psi)^\prime=U\star D_\mu\star\psi
\label{2.5}
\end{equation}

\noindent
where $U(\alpha)$ is related to the noncommutative $U(1)$ symmetry
group, namely,

\begin{equation}
U(\alpha)=\exp\star(i\alpha)
\label{2.6}
\end{equation}

\noindent
Consequently,

\begin{equation}
U^{-1}(\alpha)=\exp\star(-\,i\alpha)
\label{2.7}
\end{equation}

\noindent
Actually, by using the definition of the $\star$-product given by
(\ref{1.1}), we verify the relation

\begin{equation}
U\star U^{-1}=1
\label{2.8}
\end{equation}

\noindent
From (\ref{2.5}) and (\ref{2.8}) one obtains the gauge transformation
for $A_\mu$

\begin{equation}
A_\mu^\prime=\frac{i}{e}\,U\star D_\mu\star U^{-1}
\label{2.9}
\end{equation}

\noindent
and it is not difficult to verify that indeed (\ref{2.1}) is invariant
under (\ref{2.9}).

\medskip
Let us now consider the massive case. The action reads

\begin{equation}
S=\int d^4x\,\left(-\,\frac{1}{4}F_{\mu\nu}F^{\mu\nu}
+\frac{1}{2}\,m^2A_\mu A^\mu\right)
\label{2.10}
\end{equation}

\noindent
By virtue of the mass term, this theory is not gauge invariant. We
know that the gauge invariance can be attained by means of
Stuckelberg compensating fields \cite{Amorim2}, whose general idea
consists in extending the configuration space by the introduction of
independent noncommutative $U(1)$ group elements $g$ in the same
representation of $U(\alpha)$ introduced above. The action
(\ref{2.10}) is then rewritten in terms of modified gauge fields $\bar
A_\mu$, namely

\begin{equation}
\bar A_\mu=g\star A_\mu\star g^{-1}
+ig\star\partial_\mu g^{-1}
\label{2.11}
\end{equation}

\noindent
Since $\bar F_{\mu\nu}=g\star F_{\mu\nu}\star g^{-1}$, the
kinectical term in (\ref{2.10}) is not modified, due to the
properties of the Moyal product. It is easy to proof, however, that
the modified mass term becomes invariant under the infinitesimal gauge
transformations

\begin{eqnarray}
\delta A_\mu&=&D_\mu\star\alpha
\nonumber\\
&=&\partial_\mu\alpha-ie[A_\mu,\alpha]
\nonumber\\
\delta g&=&-ieg\star\alpha
\label{2.12}
\end{eqnarray}

\noindent
which lead to $\delta\bar A_\mu=0$. Of course the old theory is
recovered in the unitary gauge $g=1$.

\section{Embedding of massive noncommutative U(1) theory}
\renewcommand{\theequation}{3.\arabic{equation}}
\setcounter{equation}{0}

According to (\ref{2.10}), the Lagrangian density corresponding to the
massive noncommutative $U(1)$ theory is given by

\begin{equation}
{\cal L}=-\,\frac{1}{4}\,F_{\mu\nu}F^{\mu\nu}
+\frac{1}{2}\,m^2\,A_\mu A^\mu
\label{3.1}
\end{equation}

\noindent
The Moyal product does not explicitly appear in the equation above,
but it is implicit into the expression of $F_{\mu\nu}$ given by
(\ref{2.4}). This means that we have to be careful in calculating
canonical momenta starting from the Lagrangian density (\ref{3.1}),
because it contains an infinity number of time derivatives. Even
though
one can calculate momenta in theories with an infinite number of time
derivatives \cite{Barc4}, this would lead to causality and unitarity
problems. These can be circumvented in the Moyal space by taking
$\theta^{0i}=0$ \cite{Gomis}. Hence, the $\star$-product of the gauge
fields into the stress tensor $F_{\mu\nu}$ turns to be

\begin{equation}
A_\mu(x)\star A_\nu(x)=\exp\left(\frac{i}{2}\theta^{ij}
\partial_i^x\partial_j^y\right)A_\mu(x)A_\nu(y)\vert_{x=y}
\label{3.2}
\end{equation}

\noindent
Now, one can calculate the canonical momentum conjugate to $A_\mu$ in
a direct way,

\begin{equation}
\pi^\mu=\frac{\partial{\cal L}}{\partial\dot A_\mu}
=F^{\mu0}
\label{3.3}
\end{equation}

\noindent
We are using the following convention for the flat metric
$\eta^{\mu\nu}={\rm diag.} \,(1,-1,-1,-1)$. We observe that $\pi^0$ is
a primary constraint \cite{Dirac}, that we denote by $T_1$,

\begin{equation}
T_1=\pi^0
\label{3.4}
\end{equation}

\noindent
In order to look for secondary constraints, we construct the canonical
Hamiltonian

\begin{eqnarray}
H_c&=&\int d^3x\,\bigl(\pi^\mu\star\dot A_\mu-{\cal L}\bigr)
\nonumber\\
&=&\int d^3x\,\bigl(\pi^\mu\dot A_\mu-{\cal L}\bigr)
\label{3.5}
\end{eqnarray}

\noindent
where the last step was only possible inside an integration over
$d^3x$ because we are taking $\theta^{0i}=0$. Using the expression
(\ref{3.1}) we obtain for the total Hamiltonian

\begin{eqnarray}
H_T&=&H_c+\int d^3x\,\lambda T_1
\nonumber\\
&=&\int d^3x\,\Bigl(-\frac{1}{2}\pi^i\pi_i-A_0\,\partial_i\pi^i
+ie\,\pi^i[A_0,A_i]
\nonumber\\
&&+\frac{1}{4}F^{ij}F_{ij}
-\frac{1}{2}m^2A^\mu A_\mu+\lambda T_1\Bigr)
\label{3.6}
\end{eqnarray}

\noindent
The term $\pi^0\dot A_0$ of the canonical Hamiltonian density was
absorbed into $\lambda T_1$ by a redefinition of the Lagrange
multiplier $\lambda$.

\medskip
To calculate the consistency condition for the constraint $\pi^0=0$,
we have to make the Poisson brackets between $\pi^0$ and $H_T$ equal
to zero. Since Poisson brackets are taken at equal time and we are
considering $\theta^{0i}=0$, the Poisson bracket definition between
two generic quantities $F$ and $G$ is given in terms of the usual
product, (but we mention that the Dirac brackets are not) i.e.

\begin{eqnarray}
&&\{F(x),G(y)\}_{x_0=y_0}
\nonumber\\
&&\phantom{\{F}
=\int d^3z\,
\left(\frac{\delta F(\vec x)}{\delta A_\mu(\vec z)}
\frac{\delta G(\vec y)}{\delta \pi^\mu(\vec z)}
-\frac{\delta G(\vec y)}{\delta A_\mu(\vec z)}
\frac{\delta F(\vec x)}{\delta \pi^\mu(\vec z)}\right)
\label{3.7}
\end{eqnarray}

\noindent
The consistency condition for $T_1$ is

\end{multicols}
\renewcommand{\theequation}{3.\arabic{equation}}
\Lrule

\begin{eqnarray}
\{\pi_0(x),H_T\}&=&-\,\frac{\delta}{\delta A_0(x)}\,H_T
\nonumber\\
&=&\partial_i\pi^i(x)+m^2A_0(x)
-ie\,\int d^3y\,\pi^i(y)\Bigl(\delta^3(x-y)\star A_i(y)
-A_i(y)\star\delta^3(x-y)\Bigr)
\nonumber\\
&=&\partial_i\pi^i(x)+m^2A_0(x)
-ie\,\int d^3y\,\pi^i(y)\star\delta^3(x-y)\star A_i(y)
+ie\,\int d^3y\,\pi^i(y)\star A_i(y)\star\delta^3(x-y)
\nonumber\\
&=&\partial_i\pi^i(x)+m^2A_0(x)
-ie\,\int d^3y\,A_i(y)\star\pi^i(y)\star\delta^3(x-y)
+ie\,\int d^3y\,\pi^i(y)\star A_i(y)\star\delta^3(x-y)
\nonumber\\
&=&\partial_i\pi^i(x)+m^2A_0(x)
-ie\,\int d^3y\,[A_i(y),\pi^i(y)]\,\delta^3(x-y)+m^2A_0(x)
\nonumber\\
&=&\partial_i\pi^i(x)+m^2A_0(x)-ie\,[A_i(x),\pi^i(x)]
\nonumber\\
&=&D_i\star\pi^i(x)+m^2A_0(x)
\label{3.8}
\end{eqnarray}

\Rrule
\begin{multicols}{2}

\noindent
where it is understood the equal time condition for the Poisson
brackets. In the last steps it was used the identities
(\ref{A.1})-(\ref{A.3}).

\medskip
The result given by (\ref{3.8}) is the constraint $T_2(x)$ and there
are no more constraints. This is verified by showing that constraints
$T_1$ and $T_2$ satisfy a non-involutive algebra. In fact

\begin{eqnarray}
\{T_1(x),T_1(y)\}&=&0
\label{3.9}\\
\{T_1(x),T_2(y)\}&=&-\,m^2\delta^3(x-y)
\label{3.10}\\
\{T_2(x),T_2(y)\}&=&ie\,[D_i\star\pi^i(x),\delta^3(x-y)]
\label{3.11}
\end{eqnarray}

\noindent
The Eqs. (\ref{3.9}) and (\ref{3.10}) are directly obtained without
effort. See Appendix B for the obtainment of (\ref{3.11}).

\medskip
Let us now start to use the BFFT method. From expressions
(\ref{3.9})-(\ref{3.11}) we identify the quantities (we are going to
use the same notation as the one of reference \cite{Barc2})

\begin{eqnarray}
&&\Delta_{11}(x,y)=0
\nonumber\\
&&\Delta_{12}(x,y)=-\,m^2\delta^3(x-y)=-\,\Delta_{21}(x,y)
\nonumber\\
&&\Delta_{22}(x,y)=ie\,[D_i\star\pi^i(x),\delta^3(x-y)]
\label{3.12}
\end{eqnarray}

\noindent
We extend the phase space by introducing one new auxiliary field for
each (second-class) constraint. We take them as conjugate
(unconstrained) fields, i.e.

\begin{eqnarray}
&&\{\eta^1(x),\eta^2(y)\}=\delta^3(x-y)
\nonumber\\
&&\{\eta^1(x),\eta^1(y)\}=0=\{\eta^2(x),\eta^2(y)\}
\label{3.13}
\end{eqnarray}

\noindent
which permit us to identify the symplectic matrix $(\omega^{ab})=
(\{\eta^a,\eta^b\})$

\begin{equation}
\Bigl(\omega^{ab}(x,y)\Bigr)
=\left(\begin{array}{cc}
0&1\\
-\,1&0
\end{array}\right)\,\delta^3(x-y)
\label{3.14}
\end{equation}

The first correction of the constraints,

\begin{equation}
T^{(1)}_a(x)=\int d^3y\,X_{ab}(x,y)\star\eta^b(y)
\label{3.15}
\end{equation}

is achieved by solving the equation

\end{multicols}
\renewcommand{\theequation}{3.\arabic{equation}}
\Lrule

\begin{eqnarray}
&&\Delta_{ab}(x,y)+\int d^3z\,d^3z^\prime\,
X_{ac}(x,z)\star\omega^{cd}(z,z^\prime)\star X_{bd}(y,z^\prime)=0
\nonumber\\
\Longrightarrow\,\,\,
&&\Delta_{ab}(x,y)+\int d^3z\,d^3z^\prime\,
X_{ac}(x,z)\,\omega^{cd}(z,z^\prime)\,X_{bd}(y,z^\prime)=0
\label{3.16}
\end{eqnarray}

\noindent
Considering (\ref{3.12}), (\ref{3.14}), and (\ref{3.16}), we have

\begin{eqnarray}
a=1,\,b=1:\hspace{.5cm}
&&\int d^3z\,\Bigl(X_{11}(x,z)X_{12}(y,z)
-X_{12}(x,z)X_{11}(y,z)\Bigr)=0
\label{3.17}\\
a=1,\,b=2:\hspace{.5cm}
&&\int d^3z\,\Bigl(X_{11}(x,z)X_{22}(y,z)
-X_{12}(x,z)X_{21}(y,z)\Bigr)=m^2\,\delta^3(x-y)
\label{3.18}\\
a=2,\,b=2:\hspace{.5cm}
&&\int d^3z\,\Bigl(X_{21}(x,z)X_{22}(y,z)
-X_{22}(x,z)X_{21}(y,z)\Bigr)=-ie\,[D_i\star\pi^i(x),\delta^3(x-y)]
\label{3.19}
\end{eqnarray}

\Rrule
\begin{multicols}{2}

\noindent
The set given by (\ref{3.17}) - (\ref{3.19}) has more unknowns
than the number of equations. Let us choose

\begin{eqnarray}
X_{11}(x,y)&=&0
\nonumber\\
X_{21}(x,y)&=&\delta^3(x-y)
\label{3.20}
\end{eqnarray}

\noindent
We thus see that (\ref{3.17}) is automatically verified. Eqs.
(\ref{3.18}) and (\ref{3.19}) lead to

\begin{equation}
X_{12}(x,y)=-\,m^2\,\delta^3(x-y)
\label{3.21}
\end{equation}

\begin{equation}
X_{22}(x,y)-X_{22}(y,x)=ie\,[D_i\star\pi^i(x),\delta^3(x-y)]
\label{3.22}
\end{equation}

\noindent
Looking at both sides of Eq. (\ref{3.22}) we are tempted to identify
$X_{22}(x,y)$ with $ie\,D_i\star\pi^i(x)\star\delta^3(x-y)$. However,
we also observe that only the antisymmetric part of $X_{22}(x,y)$
contributes. Keeping just this part we have

\begin{eqnarray}
&&X_{22}(x,y)
\nonumber\\
&&\phantom{X}
=\frac{ie}{2}\Bigl(D_i\star\pi^i(x)\star\delta^3(x-y)
-D_i\star\pi^i(y)\star\delta^3(x-y)\Bigr)
\nonumber\\
&&\phantom{X}
=\frac{ie}{2}\Bigl(D_i\star\pi^i(x)\star\delta^3(x-y)
-\delta^3(x-y)\star D_i\star\pi^i(x)\Bigr)
\nonumber\\
&&\phantom{X}
=\frac{ie}{2}\,[D_i\star\pi^i(x),\delta^3(x-y)]
\label{3.23}
\end{eqnarray}

\noindent
In the first to the second step it was used the identity (\ref{A.4}).
The combination of (\ref{3.15}), (\ref{3.20}), (\ref{3.21}), and
(\ref{3.23}) leads to the first correction of the constraints

\begin{eqnarray}
T_1^{(1)}(x)&=&-m^2\eta^2(x)
\label{3.24}\\
T_2^{(1)}(x)&=&\eta^1(x)
-\frac{ie}{2}[\eta^2(x),D_i\star\pi^i(x)]
\label{3.25}
\end{eqnarray}

\noindent
See Appendix C for the obtainment of $T_2^{(1)}(x)$

\medskip
The next task is to evaluate the second correction $T_a^{(2)}$. This
can be achieved from the general expression

\begin{eqnarray}
&&\{T_a,T_b^{(1)}\}+\{T_a^{(1)},T_b\}
+\{T_a^{(1)},T_b^{(2)}\}_{(\eta)}
\nonumber\\
&&\phantom{\{T_a,T_b^{(1)}\}+\{T_a^{(1)},T_b\}}
+\{T_a^{(2)},T_b^{(1)}\}_{(\eta)}=0
\label{3.26}
\end{eqnarray}

\noindent
Using the expressions for the constraints $T_a$, given by (\ref{3.4})
and (\ref{3.8}), as well as (\ref{3.24}) and (\ref{3.25}), we obtain

\begin{eqnarray}
&&a=1,\,b=1:
\nonumber\\
&&\phantom{a=}
\{\eta^2,T_1^{(2)}\}_{(\eta)}+\{T_1^{(2)},\eta^2\}_{(\eta)}=0
\nonumber\\
&&\label{3.27}\\
&&a=1,\,b=2:
\nonumber\\
&&\phantom{a=}
m^2\{\eta^2,T_2^{(2)}\}_{(\eta)}
-\{T_1^{(2)},\eta^1-\frac{ie}{2}[\eta^2,D_i\star\pi^i]\}_{(\eta)}=0
\nonumber\\
&&\label{3.28}\\
&&a=2,\,b=1:
\nonumber\\
&&\phantom{a}
\{\eta^1-\frac{ie}{2}[\eta^2,D_i\star\pi^i],T_1^{(2)}\}_{(\eta)}
-m^2\{T_2^{(2)},\eta^2\}_{(\eta)}=0
\nonumber\\
&&\label{3.29}\\
&&a=2,\,b=2:
\nonumber\\
&&\phantom{a}
\frac{ie}{2}\{D_i\star\pi^i,[\eta^2,D_j\star\pi^j]\}
+\frac{ie}{2}\{[\eta^2,D_j\star\pi^j],D_i\star\pi^i\}
\nonumber\\
&&\phantom{\frac{ie}{2}\{D_i\star\pi^i,[\eta^2,}
-\{\eta^1-\frac{ie}{2}[\eta^2,D_i\star\pi^i],T_2^{(2)}\}_{(\eta)}
\nonumber\\
&&\phantom{\frac{ie}{2}\{D_i\star\pi^i,[\eta^2,}
-\{T^{(2)},\eta^1-\frac{ie}{2}[\eta^2,D_i\star\pi^i]\}_{(\eta)}=0
\nonumber\\
&&\label{3.30}
\end{eqnarray}

\noindent
The solution of (\ref{3.30}) is (see Appendix D)

\begin{equation}
T_2^{(2)}(x)=\frac{(ie)^2}{6}\,
[\eta^2(x),[\eta^2(x),D_i\star\pi^i(x)]
\label{3.31}
\end{equation}

Looking at the remaining equations (\ref{3.27})-(\ref{3.29}) we
observe that the solution we have obtained for $T_2^{(2)}$ does not
give any contribution to those equations. So, if one takes
$T_1^{(2)}=0$ into those remaining equations, they will be
automatically verified.

\medskip
We can go on and calculate other corrections for the constrains. The
procedure is similar to this last calculation of $T_a^{(2)}$. We just
write down that the general solution for the embedding procedure of
the massive noncommutative U(1) theory is given by

\begin{eqnarray}
\tilde T_1&=&\pi_0-m^2\eta^2
\label{3.32}\\
\tilde T_2&=&D_i\star\pi^i+m^2A_0
+\eta^1-\frac{ie}{2!}[\eta^2,D_i\star\pi^i]
\nonumber\\
&&\phantom{D_i\star\pi^i}
+\frac{(-ie)^2}{3!}[\eta^2,[\eta^2,D_i\star\pi^i]]
\nonumber\\
&&\phantom{D_i\star\pi^i}
+\frac{(-ie)^3}{4!}[\eta^2,[\eta^2,[\eta^2,D_i\star\pi^i]]]
\nonumber\\
&&\phantom{D_i\star\pi^i}
+\dots
\label{3.33}
\end{eqnarray}

\section{Time evolution and gauge invariance}
\renewcommand{\theequation}{4.\arabic{equation}}
\setcounter{equation}{0}

So far we have succeeded in finding the converted constraints $\tilde
T_1$ and $\tilde T_2$, which are in involution, as can be verified. To
get a complete description of the dynamics, it is also necessary to
obtain an involutive Hamiltonian. This can be done by using directly
the BFFT procedure. A simpler approach, however, is the one used in
Ref.\cite{Barc2} where modified phase space variables $\tilde A_\mu$
and $\tilde \pi_\mu$ are constructed in such a way that they are
involutive with the converted constraints  $\tilde T_1$ and $\tilde
T_2$ and at the same time recover the original phase space variables
$A_\mu$ and $\Pi_\mu$ in the unitary gauge. With the aid of $\tilde
A_\mu$ and $\tilde \pi_\mu$, we then directly obtain the proper
involutive Hamiltonian by the rule $\tilde H(A_\mu,\pi_\mu,\eta^a)=
H(\tilde A_\mu,\tilde\pi_\mu)$, where $H(A_\mu,\pi_\mu)$  given by
(\ref{3.6}). As the calculations are long and similar to those
performed in the previous sections, we just display the resulting
expressions for $\tilde A_\mu$ and $\tilde\pi_\mu$:

\begin{eqnarray}
\tilde A_0&=&A_0
+\frac{1}{m^2}\eta^1
+\frac{ie}{2m^2}\,[\eta^2,D_i\star\pi^i]
\nonumber\\
&&-\,\frac{(ie)^2}{3m^2}\,[\eta^2,[\eta^2,D_i\star\pi^i]]+\cdots
\nonumber\\
&&-\,\frac{n}{(n+1)!}\frac{(-ie)^n}{m^2}\,[\eta^2,[\eta^2,\dots,
[\eta^2,D_i\star\pi^i]]\dots]
\nonumber\\
&&\label{4.1}\\
\tilde A_i&=&A_i
-D_i\star\eta^2
+\frac{ie}{2}\,[\eta^2,D_i\star\eta^2]
\nonumber\\
&&-\,\frac{(-ie)^2}{3!}\,[\eta^2,[\eta^2,D_i\star\eta^2]]+\cdots
\label{4.2}\\
\tilde\pi_0&=&\pi_0
\label{4.3}\\
\tilde\pi_i&=&\pi_i
-ie\,[\eta^2,\pi_i]
+\frac{(-ie)^2}{2}\,[\eta^2,[\eta^2,\pi_i]]
\nonumber\\
&&+\,\frac{(-ie)^3}{3!}\,[\eta^2,[\eta^2,[\eta^2,\pi_i]]]+\cdots
\label{4.4}
\end{eqnarray}

\noindent
Since $\tilde T_1$ and $\tilde T_2$ are involutive with respect to
$\tilde H$, they are consistently conserved. Also, as they are first
class, they generate the gauge transformations inside the Hamiltonian
formalism. Let $\omega^a$, $a=1,2$ be arbitrary infinitesimal
functions. Then $G[\omega^a]=\int d^3x\,(\omega^a\tilde T_a)$
generates infinitesimal gauge transformations in the phase space
variables $y^\alpha$ given by $\delta y^\alpha=\{y^\alpha, G\}$. It
is
then possible to show \cite{Henneaux} that the first order action

\begin{equation}
S_{f0}=\int d^4x\,(\pi^\mu \dot A_\mu+\eta^2\dot\eta^1
-\tilde H-\lambda^a\tilde T_a)
\label{4.5}
\end{equation}

\noindent
is gauge invariant since the Hamiltonian and the constraints satisfy
an Abelian algebra, once $\delta\lambda^a=\dot\omega^a$. It is not
difficult to verify that the  gauge transformations of the phase space
coordinates are given by

\begin{eqnarray}
\delta A_0&=&-\,\omega_1
\nonumber\\
\delta A_i&=&-\,D_i\star\biggl(\omega_2
-\frac{ie}{2!}\,[\omega_2,\eta^2]
+\frac{(-ie)^2}{3!}[[\omega^2,\eta^2],\eta^2]
\nonumber\\
&&+\,\frac{(-ie)^3}{4!}[[[\omega^2,\eta^2],\eta^2],\eta^2]
+\cdots\biggr)
\nonumber\\
\delta\pi_0&=&-\,m^2\omega_2
\nonumber\\
\delta\pi_i&=&ie\,[\pi_i,\omega_2]
-\,\frac{(ie)^2 }{2!}\,[\eta^2,[\pi_i,\omega_2]]
\nonumber\\
&&+\,\frac{(ie)^3 }{3!}\,[\eta^2,[\eta^2,[\pi_i,\omega_2]]]+\cdots
\nonumber\\
\delta\eta^1&=&-\,m^2\omega_1
-\frac{ie}{2!}\,[D_i\star\pi^i,\omega_2]
\nonumber\\
&&-\,\frac{(ie)^2}{3!}\,[D_i\star\pi^i,[\eta^2,\omega_2]]
\nonumber\\
&&-\,\frac{(ie)^3}{4!}\,[D_i\star\pi^i,[\eta^2,[\eta^2,\omega_2]]]
-\cdots
\nonumber\\
\delta\eta^2&=&\omega_2
\label{4.6}
\end{eqnarray}

\noindent
Observe that these transformations have the proper commutative limit
but are not trivially related to the Lagrangian ones, given by
(\ref{2.12}).

\section{Conclusion}
In this paper we have studied the embedding of the massive
noncommutative $U(1)$ theory by using the BFFT Hamiltonian formalism.
Due to the peculiar mathematical structure of the algebra
related to the Moyal space, some care was necessary in order to deal
with the iterative steps of the method. For this part we emphasize the
importance of the identities (\ref{A.4}) and (\ref{A.8}). We succeeded
in finding the converted constraints and Hamiltonian, which are in
involution. They were obtained in terms of infinite series of Moyal
commutators among the BFFT variables and functions of the old phase
space. The gauge structure of the embedded theory was also displayed.

\vspace{1cm}
\noindent
{\bf Acknowledgment:} This work is supported in part by Conselho
Nacional de Desenvolvimento Cient\'{\i}fico e Tecnol\'ogico
- CNPq (Brazilian Research agency) with the support of PRONEX
66.2002/1998-9.

\section*{Appendix A}
\section*{Some identities related to the star product}
\renewcommand{\theequation}{A.\arabic{equation}}
\setcounter{equation}{0}

In this Appendix we list the main identities used in the paper

\begin{equation}
\int d^4x\,\phi_1\star\phi_2=\int d^4x\,\phi_1\phi_2
=\int d^4x\,\phi_2\star\phi_1
\label{A.1}
\end{equation}

\begin{equation}
\bigl(\phi_1\star\phi_2\bigr)\star\phi_3
=\phi_1\star\bigl(\phi_2\star\phi_3\bigr)
=\phi_1\star\phi_2\star\phi_3
\label{A.2}
\end{equation}

\begin{eqnarray}
\int d^4x \,\phi_1\star\phi_2\star\phi_3
&=&\int d^4x\,\phi_2\star\phi_3\star\phi_1
\nonumber\\
&=&\int d^4x\,\phi_3\star\phi_1\star\phi_2
\label{A.3}
\end{eqnarray}

\begin{equation}
\phi(x)\star\delta(x-y)=\delta(x-y)\star\phi(y)
\label{A.4}
\end{equation}

\begin{equation}
D_\mu\star(\phi_1\star\phi_2)=(D_\mu\star\phi_1)\star\phi_2
+\phi_1\star(D_\mu\star\phi_2)
\label{A.5}
\end{equation}

\begin{equation}
D_\mu^x\star\delta(x-y)=-\,D_\mu^y\star\delta(x-y)
\label{A.6}
\end{equation}

\begin{equation}
[\phi_1,[\phi_2,\phi_3]]+[\phi_2,[\phi_3,\phi_1]]
+[\phi_3,[\phi_1,\phi_2]]=0
\label{A.7}
\end{equation}

\begin{equation}
[\phi_1(x),[\phi_2(x),\delta(x-y)]]
=[\phi_2(y),[\phi_1(y),\delta(x-y)]]
\label{A.8}
\end{equation}

\end{multicols}
\Lrule

\section*{Appendix B}
\section*{Obtainment of (\ref{3.11})}
\renewcommand{\theequation}{B.\arabic{equation}}
\setcounter{equation}{0}

\begin{eqnarray}
\{T_2(x),T_2(y)\}&=&\{D_i\star\pi^i(x),D_j\star\pi^j(y)\}
\nonumber\\
&=&\int d^3z\,\biggl(
\frac{\delta^3}{\delta^3 A_k(z)}\Bigl(D_i\star\pi^i(x)\Bigr)
\frac{\delta^3}{\delta^3 \pi^k(z)}\Bigl(D_j\star\pi^j(y)\Bigr)
-\frac{\delta^3}{\delta^3 A_k(z)}\Bigl(D_j\star\pi^j(y)\Bigr)
\frac{\delta^3}{\delta^3 \pi^k(z)}\Bigl(D_i\star\pi^i(x)\Bigr)
\biggr)
\nonumber\\
&=&-ie\int d^3z\Bigl(
[\delta^3(x-z),\pi^k(x)]\,D_k^y\star\delta^3(y-z)
-[\delta^3(y-z),\pi^k(y)]\,D_k^x\star\delta^3(x-z)
\Bigr)
\nonumber\\
&=&-ie\,D_k^y\star\int d^3z\,[\delta^3(x-z),\pi^k(x)]\,\delta^3(y-z)
+ie\,D_k^x\star\int d^3z\,[\delta^3(y-z),\pi^k(y)]\,\delta^3(x-z)
\nonumber\\
&=&-ie\,D_k^y\star\Bigl([\delta^3(x-y),\pi^k(x)]\Bigr)
+ie\,D_k^x\star\Bigl([\delta^3(y-x),\pi^k(y)]\Bigr)
\nonumber\\
&=&-ie\,[D_k^y\star\delta^3(x-y),\pi^k(x)]
+ie\,D_k^x\star\Bigl([\pi^k(x),\delta^3(x-y)]\Bigr)
\label{B.1}
\end{eqnarray}

\noindent
where it was used the identity (\ref{A.4}) for the second term of the
last step above. Considering (\ref{A.5}) and (\ref{A.6}) we have

\begin{eqnarray}
\{T_2(x),T_2(y)\}&=&ie\,[D_k^x\star\delta^3(x-y),\pi^k(x)]
+ie\,[D_k\star\pi^k(x),\delta^3(x-y)]
+ie\,[\pi^k(x),D_k^x\star\delta^3(x-y)]
\nonumber\\
&=&ie\,[D_k\star\pi^k(x),\delta^3(x-y)]
\label{B.2}
\end{eqnarray}

\begin{multicols}{2}

\section*{Appendix C}
\section*{Obtainment of $T_2^{(1)}$}
\renewcommand{\theequation}{C.\arabic{equation}}
\setcounter{equation}{0}

Making the replacement of (\ref{3.20}) and (\ref{3.23}) into
(\ref{3.15}) we get

\begin{eqnarray}
T_2^{(1)}&=&\int d^3y\,\delta^3(x-y)\star\eta^1(y)
\nonumber\\
&&+\frac{ie}{2}\int d^3y\,[D_i\star\pi(x),\delta^3(x-y)]\star\eta^2(y)
\nonumber\\
&=&\eta^1(x)
+\frac{ie}{2}\int d^3y\,
D_i\star\pi^i(x)\star\delta^3(x-y)\star\eta^2(y)
\nonumber\\
&&-\frac{ie}{2}\int d^3y\,
\delta^3(x-y)\star D_i\star\pi^i(x)\star\eta^2(y)
\label{C.1}
\end{eqnarray}

\noindent
In the triple Moyal product above involving $D_i\star\pi^i(x)$,
$\eta^2(y)$ and $\delta^3(x-y)$ one cannot use the rule given by
(\ref{A.3}) because these products do not refer to the same
variable. Let us then conveniently developed each integral in a
separate way.

\begin{eqnarray}
&&\int d^3y\,D_i\star\pi^i(x)\star\delta^3(x-y)\star\eta^2(y)
\nonumber\\
&&\phantom{\int d^3y\,}
=D_i\star\pi^i(x)\star\int d^3y\,\delta^3(x-y)\star\eta^2(y)
\nonumber\\
&&\phantom{\int d^3y\,}
=D_i\star\pi^i(x)\star\int d^3y\,\delta^3(x-y)\,\eta^2(y)
\nonumber\\
&&\phantom{\int d^3y\,}
=D_i\star\pi^i(x)\star\eta^2(x)
\label{C.2}
\end{eqnarray}

\noindent
For the second integral, we have

\begin{eqnarray}
&&\int d^3y\,\delta^3(x-y)\star D_i\star\pi^i(x)\star\eta^2(y)
\nonumber\\
&&\phantom{\int d^3y\,}
=\int d^3y\,\delta^3(x-y)\star D_i\star\pi^i(x)\,\eta^2(y)
\nonumber\\
&&\phantom{\int d^3y\,}
=\left(\int d^3y\,\eta^2(y)\delta^3(x-y)\right)\star D_i\star\pi^i(x)
\nonumber\\
&&\phantom{\int d^3y\,}
=\eta^2(x)\star D_i\star\pi^i(x)
\label{C.3}
\end{eqnarray}

The replacement of (\ref{C.2}) and (\ref{C.3}) into (\ref{C.1}) leads
to $T_2^{(1)}(x)$  given by (\ref{3.25}).

\end{multicols}
\Lrule

\section*{Appendix D}
\section*{Solution of Eq. (\ref{3.30})}
\renewcommand{\theequation}{D.\arabic{equation}}
\setcounter{equation}{0}

Let us develop the two first terms of (\ref{3.30}) in a separate way.
For the first term, we have

\begin{eqnarray}
-\frac{ie}{2}\{D_i\star\pi^i,[\eta^2,D_j\star\pi^j]\}
&=&-\frac{ie}{2}\{D_i\star\pi^i(x),\eta^2(y)\star D_j\star\pi^j(y)
-D_j\star\pi^j(y)\star\eta^2(y)\}
\nonumber\\
&=&-\frac{ie}{2}\eta^2(y)\star\{D_i\star\pi^i(x),D_j\star\pi^j(y)\}
+\frac{ie}{2}\{D_i\star\pi^i(x),D_j\star\pi^j(y)\}\star\eta^2(y)
\nonumber\\
&=&\frac{ie}{2}[\{D_i\star\pi^i(x),D_j\star\pi^j(y)\},\eta^2(y)]
\label{D.1}
\end{eqnarray}

\noindent
Using (\ref{A.4}), (\ref{A.8}), and (\ref{3.11}), we obtain

\begin{eqnarray}
-\frac{ie}{2}\{D_i\star\pi^i,[\eta^2,D_j\star\pi^j]\}
&=&\frac{(ie)^2}{2}[[D_i\star\pi^i(x),\delta^3(x-y)],\eta^2(y)]
\nonumber\\
&=&\frac{(ie)^2}{2}[[\delta^3(x-y),D_i\star\pi^i(y)],\eta^2(y)]
\nonumber\\
&=&\frac{(ie)^2}{2}[[\delta^3(x-y),\eta^2(x)],D_i\star\pi^i(x)]
\label{D.2}
\end{eqnarray}

\noindent
Developing the second term in a similar way, the result is

\begin{equation}
-\frac{ie}{2}\{[\eta^2,D_j\star\pi^j],D_i\star\pi^i\}
=\frac{(ie)^2}{2}[[D_i\star\pi^i(x),\delta^3(x-y)],\eta^2(x)]
\label{D.3}
\end{equation}

\noindent
Let us introduce these two terms into expression (\ref{3.30}),

\begin{eqnarray}
&&\{T_2^{(2)},\eta^1-\frac{ie}{2}[\eta^2,D_i\star\pi^i]\}_{(\eta)}
+\{\eta^1-\frac{ie}{2}[\eta^2,D_i\star\pi^i],T_2^{(2)}\}_{(\eta)}
\nonumber\\
&&\phantom{\{T_2^{(2)},\eta^1}
=-\frac{(ie)^2}{2}\biggl([[\delta^3(x-y),\eta^2(x)],D_i\star\pi^i(x)]
+[[D_i\star\pi^i(x),\delta^3(x-y)],\eta^2(x)]\biggr)
\nonumber\\
&&\phantom{\{T_2^{(2)},\eta^1}
=\frac{(ie)^2}{2}[[\eta^2(x),D_i\star\pi^i(x)],\delta^3(x-y)]
\label{D.4}
\end{eqnarray}

\noindent
where it was used the Jacobi identity (\ref{A.7}).

\medskip
Looking at both sides of (\ref{D.3}) we observe that the only
possibility of solution is considering $T_2^{(2)}$ quadratic in
$\eta^2$ (it could not depend on $\eta^1\eta^2$ or $\eta^1\eta^1$
because this would lead to terms in $\eta^1$ in the left side what
would be inconsistent with the right side). We infer that an
appropriate form for $T_2^{(2)}$ should be

\begin{equation}
T_2^{(2)}(x)=\frac{(ie)^2}{6}
\,[\eta^2(x),[\eta^2(x),D_i\star\pi^i(x)]]
\label{D.5}
\end{equation}

\noindent
where $(ie)^2/6$ is a convenient factor to achieve the solution of
(\ref{3.30}). Actually, developing the left side of (\ref{D.4}) with
$T_2^{(2)}$ given by (\ref{D.5}), we get

\begin{eqnarray}
&&\frac{(ie)^2}{6}
\{[\eta^2(x),[\eta^2(x),D_i\star\pi^i(x)]],\eta^1(y)\}
+\frac{(ie)^2}{6}
\{\eta^1(x),[\eta^2(y),[\eta^2(y),D_i\star\pi^i(y)]]\}
\nonumber\\
&&\phantom{\frac{(ie)^2}{6}\{[\eta^2(x),[\eta^2(x)}
=-\frac{(ie)^2}{6}\int d^3z\,\frac{\delta^3}{\delta^3\eta^2(z)}
[\eta^2(x),[\eta^2(x),D_i\star\pi^i(x)]]\delta^3(y-z)
\nonumber\\
&&\phantom{\frac{(ie)^2}{6}\{[\eta^2(x),[\eta^2(x)=}
+\frac{(ie)^2}{6}\int d^3z\,\delta^3(x-z)
\frac{\delta^3}{\delta^3\eta^2(z)}
[\eta^2(y),[\eta^2(y),D_i\star\pi^i(y)]]
\nonumber\\
&&\phantom{\frac{(ie)^2}{6}\{[\eta^2(x),[\eta^2(x)}
=-\frac{(ie)^2}{6}
[\delta^3(x-y),[\eta^2(x),D_i\star\pi^i(x)]]
-\frac{(ie)^2}{6}
[\eta^2(x),[\delta^3(x-y),D_i\star\pi^i(x)]]
\nonumber\\
&&\phantom{\frac{(ie)^2}{6}\{[\eta^2(x),[\eta^2(x)=}
+\frac{(ie)^2}{6}
[\delta^3(x-y),[\eta^2(y),D_i\star\pi^i(y)]]
+\frac{(ie)^2}{6}
[\eta^2(y),[\delta^3(x-y),D_i\star\pi^i(y)]]
\label{D.6}
\end{eqnarray}

\noindent
From the identity (\ref{A.4}) we have that the first and the third
terms of (\ref{D.6}) are equal. The last one can be written as
$-\frac{(ie)^2}{6}[D_i\star\pi^i(x),[\eta^2(x),\delta(x-y)]]$, by
virtue of (\ref{A.8}). Using the Jacobi identity with this and the
second term of (\ref{D.6}) we obtain a term that is equal to the first
one. Hence, the four terms of (\ref{D.6}) is three times the
first term, that is precisely the right side of (\ref{D.4}) as we
want to show.

\Rrule
\begin{multicols}{2}

\end{multicols}
\end{document}